\documentclass[aps,pra,twocolumn,10pt,showpacs,amsmath,amssymb,superscriptaddress]{revtex4-1}
\usepackage{graphicx}
\usepackage{xcolor}
\usepackage{enumerate}
\usepackage{braket}
\usepackage[colorlinks]{hyperref}
\usepackage{mathbbol} %for \1, load at last
\definecolor{myblue}{rgb}{0.2,0.2,0.8}
\definecolor{myred}{rgb}{1,0.,0.3}

\DeclareMathOperator\1{\mathbb{1}}

\begin{document}

\title{Kibble--Zurek scaling of the one-dimensional Bose--Hubbard model at finite temperatures} % Force line breaks with \\

\author{Werner Weiss}
\email{werner.weiss@uni-ulm.de}
\affiliation{Institute for Complex Quantum Systems \& Center for Integrated Quantum Science and Technology, Ulm University, Albert-Einstein-Allee 11, 89069 Ulm, Germany}
\author{Matthias Gerster}
\affiliation{Institute for Complex Quantum Systems \& Center for Integrated Quantum Science and Technology, Ulm University, Albert-Einstein-Allee 11, 89069 Ulm, Germany}
\author{Daniel Jaschke}
\affiliation{Department of Physics, Colorado School of Mines, Golden, CO 80401, United States of America}
\author{Pietro Silvi}
\affiliation{Institute for Theoretical Physics, Innsbruck University, A-6020 Innsbruck, Austria}
\author{Simone Montangero}
\affiliation{Institute for Complex Quantum Systems \& Center for Integrated Quantum Science and Technology, Ulm University, Albert-Einstein-Allee 11, 89069 Ulm, Germany}
\affiliation{Theoretische Physik, Universit{\"a}t des Saarlandes, D-66123 Saarbr{\"u}cken, Germany}
\affiliation{Dipartimento di Fisica e Astronomia ``G. Galilei", Universit\`{a} degli Studi di Padova, I-35131 Padova, Italy}
% Other names...

\date{\today}

\begin{abstract}
We use tensor network methods --- Matrix Product States, Tree Tensor Networks, and Locally Purified Tensor Networks --- to simulate the one-dimensional 
Bose--Hubbard model for zero and finite temperatures in experimentally accessible regimes. We first explore the effect of thermal fluctuations on the system ground state 
by characterizing its Mott and superfluid features. Then, we study the behavior of the out-of-equilibrium dynamics induced by quenches of the hopping parameter. 
We confirm a Kibble--Zurek scaling for zero temperature and characterize the finite temperature behavior, which we explain by means of a simple argument.
\end{abstract}

					% PACS, the Physics and Astronomy Classification Scheme.
%\keywords{Suggested keywords}			%U se showkeys class option if keyword display desired

\maketitle

\section{Introduction} 

Ultracold quantum gases in optical lattices offer the possibility to explore the behavior of condensed matter systems on a controllable testbed~\cite{Jaksch1998,Jaksch2005,Lewenstein2007,Bakr2010,Bloch2008,Bloch2010,Gross2017}. This platform using interference of laser beams to create spatial standing waves is well-suited for tailoring a variety of lattice structures in three~\cite{Greiner2002} or less dimensions~\cite{Paredes2004,Stoeferle2004,Spielman2007}. Systems of bosons in an optical lattice can be described by the Bose--Hubbard model. First introduced in the 1960s by Gersch and Knollmann~\cite{Gersch1963}, it became very helpful in understanding the superfluid to Mott insulator phase transition~\cite{Fisher1989,Kuehner2000} and has been realized in a multitude of experiments (for an overview see, e.g. Ref.~\cite{Krutitsky2016}). In particular, the one-dimensional setting has been studied in great depth over the years and is characterized by rich physics, one of the reasons being the occurrence of a multicritical point with a Berezinskii--Kosterlitz--Thouless (BKT) transition~\cite{Fisher1989,Kuehner2000}.
%(see Sec. \ref{sec:dynamics})

For a long time the theoretical and numerical work on this model has concentrated on the zero-temperature limit, which  is a valid approximation for many experimental setups. Nevertheless, characterizing the impact of thermal fluctuations is an important prerequisite in order to enable a comprehensive understanding of the observed phenomena~\cite{McKay2011}. An early investigation on the influence of finite temperatures on the Bose--Hubbard model, focusing mostly on the insulating regime, has been carried out in Ref.~\cite{Gerbier2007}. This work was followed by further theoretical~\cite{Hu2009,Dalidovich2009,Hoffmann2009,Hassan2010,Mahmud2011} and experimental studies~\cite{Spielman2008,McKay2009,Jimenez-Garcia2010,Trotzky2010,Gori2016}.

In addition to the equilibrium physics of the model, the investigation of dynamical processes, arising from tuning the system's parameters, are of great interest~\cite{Kollath2005,Morsch2006,Kollath2007,Trotzky2012,Daley2012,Bernier2018}, especially towards engineering complex phases in quantum gases. An important scenario in this context are quasi-adiabatic quenches across quantum phase transitions, for which the Kibble--Zurek hypothesis~\cite{Kibble1976,Zurek1985,Zurek2005,Dziarmaga2010} offers a simple and intuitive theoretical framework, yet allowing for a quantitative understanding of the formation of defects when crossing a quantum critical point. 
%More precisely, it predicts a scaling law for the defect density as a function of the quench time when performing a linear quench across the quantum phase transition.
%However, also this framework has been developed in the context of zero temperature and the influence of thermal fluctuations is a current research topic~\cite{Yin2016} as well as studying various aspects of the quenches themselves~\cite{Dziarmaga2014,Beugnon2017,Bernier2018,Shimizu2018,Bera2018}.
The Kibble--Zurek mechanism has been tested in a plethora of theoretical and experimental settings, including the Bose--Hubbard model itself~\cite{Dziarmaga2014,Braun2015,Shimizu2018}. %\cite{Beugnon2017}
Also in this context, attempts have been made to address thermal effects~\cite{Dutta2016,Yin2016,Gao2017}.

In this work we focus on the one-dimensional Bose--Hubbard model on chains of moderate sizes in the range of current experiments. We analyze the effects of finite temperature on two types of scenarios of experimental interest: First, we characterize the properties of the system after being prepared in an initial thermal state under a given set of constant system parameters. We study to which extent the properties of the insulating and superfluid phase persist at finite temperatures, expanding on previous results~\cite{Gerbier2007}.
%We show that in the case of the considered system sizes, the properties of both the insulating and the superfluid phase persist up to certain finite temperatures, confirming the results of Gerbier~\cite{Gerbier2007}.
%Apart from the equilibrium analysis
Secondly, we explore the dynamics of the system triggered by a linear quench in the particle hopping parameter. We verify the predicted Kibble--Zurek scaling at zero temperature~\cite{Dziarmaga2014}, and then study deviations from this behavior with rising initial temperature. We propose a simple argument, capable of providing a quantitatively correct prediction of the obtained finite-temperature results.

Our analysis is based on numerical simulations using Tensor Network (TN) methods, which are well established as a powerful tool for simulating low-dimensional strongly-correlated many-body systems~\cite{Schollwock2011,Orus2014,Silvi2017}. At the core of the analysis, we employ Locally Purified Tensor Networks (LPTN)~\cite{Verstraete2004,DeLasCuevas2013,Werner2016}, a tailored variational ansatz capable of representing thermal equilibrium states, as well as to perform real-time evolution for time-dependent Hamiltonians and Lindblad master equations. Previously, this method has been successfully applied to Quantum Ising chains~\cite{Jaschke2018a}. In the zero-temperature limit, we complement our results using Matrix Product State (MPS)~\cite{Perez2007} and Tree Tensor Network (TTN)~\cite{Shi2006,Gerster2014} simulations.

The remainder of this paper is structured as follows: In Sec.~\ref{sec:static} we introduce the model and the notation, and study the properties of the system at equilibrium by characterizing the insulating (Sec.~\ref{sub:mottChar}) and superfluid (Sec.~\ref{sub:superfluidChar}) features of its thermal states.
The collected results are summarized in a finite-temperature state diagram (Sec.~\ref{sub:phaseDiag}). In Sec.~\ref{sec:dynamics} we extend our analysis to dynamical processes by quenching the system in the hopping parameter, first at zero temperature (Sec.~\ref{sub:quenchzero}), and then at finite temperatures (Sec.~\ref{sub:quenchfinite}). In Sec.~\ref{sec:conclusion} we draw our conclusions.

\section{\label{sec:static}Equilibrium properties}

We consider a 1D Bose--Hubbard lattice~\cite{Fisher1989} described by the Hamiltonian
\begin{align}
	H =& -J \sum_{j=1}^{L-1} \left( b_j^\dagger b^{}_{j+1} + \text{h.c.}  \right)  \nonumber \\ 
	&+ \frac{U}{2} \sum_{j=1}^L n_j (n_j - 1) - \mu \sum_{j=1}^L n_j \; .
	\label{eq:hamiltonian}
\end{align}
Here, $b^{}_j$ ($b^\dagger_j$) is a bosonic annihilation (creation) operator obeying $[b^{}_j, b^\dagger_{j^\prime}]=\delta_{j j^\prime}$, and $n_j=b^\dagger_j b^{}_j $ is the particle number operator on site $j$. $L$ is the length of the chain, which we assume to have open boundaries. The coupling $J$ determines the hopping strength, while~$U$ and~$\mu$ represent the on-site repulsion strength and the chemical potential, respectively. By setting $U=\hbar=k_\mathrm{B}=1$, $k_B$ being Boltzmann's constant, we fix the units of energy~$E$, time~$t$, and temperature~$T$. 

Depending on the values of the parameters $J$ and $\mu$, the ground state of $H$ at zero temperature exhibits different phase properties~\cite{Fisher1989,Kuehner1998,Kuehner2000}. Two phases emerge: In the \emph{Mott insulator} phase, which in the ($J$, $\mu$)-plane of the phase diagram appears as ``lobes''~\cite{Fisher1989} in proximity of the $J=0$ axis, the bosonic particles are localized at single lattice sites and the filling factor $\varrho=N/L$ (where $N=\langle \sum_{j=1}^L n_j \rangle$ is the total number of particles in the system) is pinned to integer values, depending on the chemical potential $\mu$. Moreover, this phase has a finite energy gap $\Delta E$ and it is incompressible, i.e.\@ $\partial \varrho / \partial \mu = 0$.
In contrast, a \emph{superfluid} phase appears for large enough~$J$, in which the bosons are delocalized over the entire lattice. In this phase, $\varrho$ is in general not integer, hence it is compressible $\partial \varrho / \partial \mu > 0$. The superfluid phase is gapless, i.e.\ \mbox{$\Delta E = 0$}, and its quasi-long-range order is expected to disappear at any finite temperature in the thermodynamic limit.

\begin{figure}
	\includegraphics[width=0.95\columnwidth]{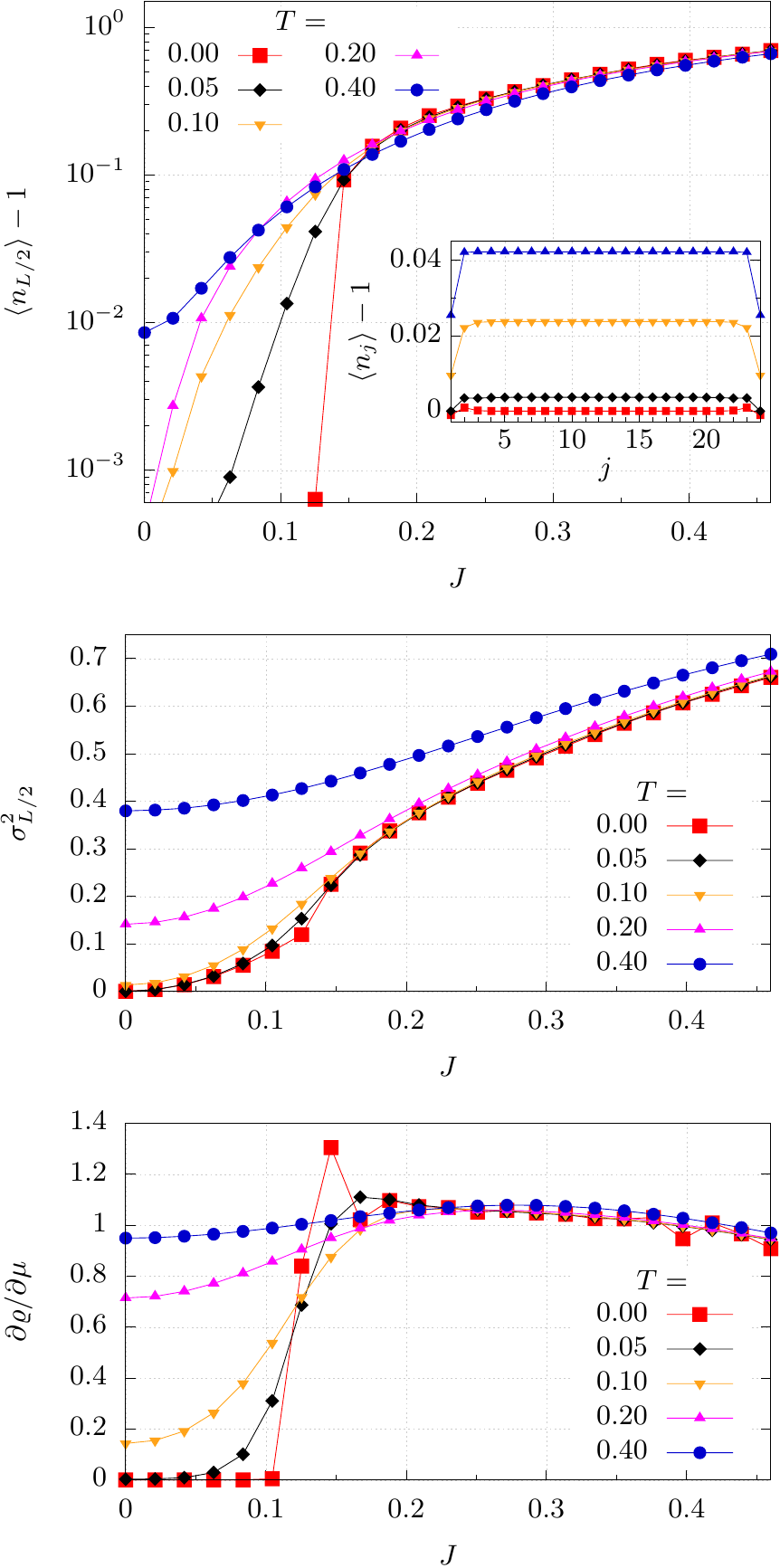}
	\caption{\label{fig:mottChar}(Color online) On-site occupation~$\langle n_{L/2} \rangle$ (top panel), variance~$\sigma^2_{L/2}$ (middle panel), and compressibility~$\partial\varrho/\partial \mu$ (bottom panel, determined via a linear fit of $\varrho(\mu)$ in the interval \mbox{$\mu \in [0.425,0.575]$}) as a function of $J$ for various temperatures~$T$, measured at the center of a chain with~$L=24$ sites for $U=1$ and $\mu=1/2$. The inset in the top panel shows the particle occupations along the whole chain, for fixed $J=0.08$.}
\end{figure}

In the remainder of this section, we characterize the equilibrium properties of the Bose--Hubbard chain at finite temperatures $T>0$. In particular, we aim to quantify to which extent the thermal equilibrium states keep their Mott- or superfuid-phase features when increasing the temperature at a finite size $L$. We perform this characterization numerically, using an LPTN ansatz state, representing the thermal many-body density matrix $\rho = e^{-\beta H} / \, \text{Tr}[e^{-\beta H}]$, with $\beta=1/T$. For zero temperature, an MPS can be used instead of an LPTN. In this sense, the LPTN extends the MPS picture, valid at $T=0$, to finite temperatures (see also Appendix~\ref{app:numerical_methods}). Clearly, the numerical treatment implies a truncation of the bosonic local Fock spaces to a finite cutoff dimension~$d$, in order to carry out the numerical simulation. The effect of this truncation is tunable, and negligible as long as high local occupation numbers are energetically suppressed, i.e.\@ as long as the parameters $J$, $\mu$, and $T$ do not become too large (compared to $U=1$). Here, we adopt up to $d=5$, which we verified to be sufficient for the parameter regime studied here, see also Appendix \ref{app:convergence}. 
% In the case of the highest simulated temperature, i.e. $T=0.5$, the population of the largest local occupation number is smaller than $7\cdot10^{-3}$.
The lengths of the simulated systems range from $L=16$ to $L=32$ sites. We target via LPTN the grand canonical ensemble density matrix, and, in what follows, we use $\mu=1/2$. Along this line in the phase diagram, the transition from the Mott insulator to the superfluid is known to be a second order quantum phase transition in the $T=0$ case~\cite{Fisher1989}, taking place at a critical hopping strength of $J_c \approx 0.13$~\cite{Kuehner2000}. Let us stress that this scenario is not to be confused with the phase transition at fixed particle filling~$\varrho \in \mathbb{N}$, which is of the BKT type~\cite{Fisher1989} and will play a role in the real-time dynamics.

\subsection{\label{sub:mottChar}Characterization of Mott insulating features}

We start by quantifying the Mott-like character of the system, as a function of both $J$ and $T$. In order to do so, we use the on-site particle occupations $\langle n_j \rangle$ and their variance
\begin{equation}
	\sigma^2_j = \langle n^2_j \rangle - \langle n_j \rangle^2 \; ,
	\label{eq:variance}
\end{equation}
as well as the compressibility $\partial\varrho/\partial \mu$.
A necessary condition for Mott insulating states are localized particles, leading to integer on-site occupation numbers $\langle n_j \rangle \in \mathbb{N}$. This behavior is accompanied by small variances $\sigma^2_j \approx 0$ and a vanishing compressibility. Specifically, the particle occupation is one (i.e.\ $\langle n_j \rangle = 1$) in the first Mott lobe which is crossed by the $\mu=1/2$ line studied here.  In contrast, outside of the Mott insulating phase, the occupation can attain any value $\langle n_j \rangle \in \mathbb{R}^+$ and the compressibility is strictly larger than zero. This finite compressibility can either be induced by thermal fluctuations, when $T$ becomes large enough to overcome the on-site repulsion, or by quantum fluctuations, even at zero temperature, when $J$ becomes large enough to favor delocalized particles.

In Fig.~\ref{fig:mottChar}, we show the numerically obtained occupation numbers, variances, and compressibilities as a function of the coupling~$J$ for various temperatures~$T \in [0, 0.4]$. Here the system size is $L=24$ sites.
%we verified that finite-size effects are negligible for these local quantities (see also Appendix~\ref{app:convergence}). 
In order to avoid boundary effects, we measure local quantities close to the center of the chain.
%We plot local quantities measured at the center of the chain which avoids boundary effects of the order $10^{-2}$ of the otherwise translationally invariant system, see inset of ~\ref{fig:mottChar}.
As shown in Fig.~\ref{fig:mottChar}, the on-site particle occupation at $T=0$ is indeed exactly $\langle n_j \rangle = 1$  in the interval $0\leq J \leq J_c$, while for $J > J_c$ a monotonous increase can be observed. This abrupt behavior is replaced by a smoother transition with rising temperature, actually reducing the range of $J$ supporting a Mott-like emergent behavior with $\langle n_j\rangle = 1$. The compressibility $\partial\varrho/\partial\mu$ exhibits a similar behavior; we remark, however, that this quantity is more prone to finite-size effects due to the involved numerical derivative (see Appendix~\ref{app:convergence}). For small temperatures $T < T^\ast \approx 0.2$, we observe that the variance is approximately insensitive to~$T$, indicating a survival of Mott-like features at least up to these temperatures. For this reason, the temperature $T^\ast$ has also been referred to as the ``melting temperature'' of the Mott insulator~\cite{Gerbier2007}.

\begin{figure}
	\includegraphics[width=0.95\columnwidth]{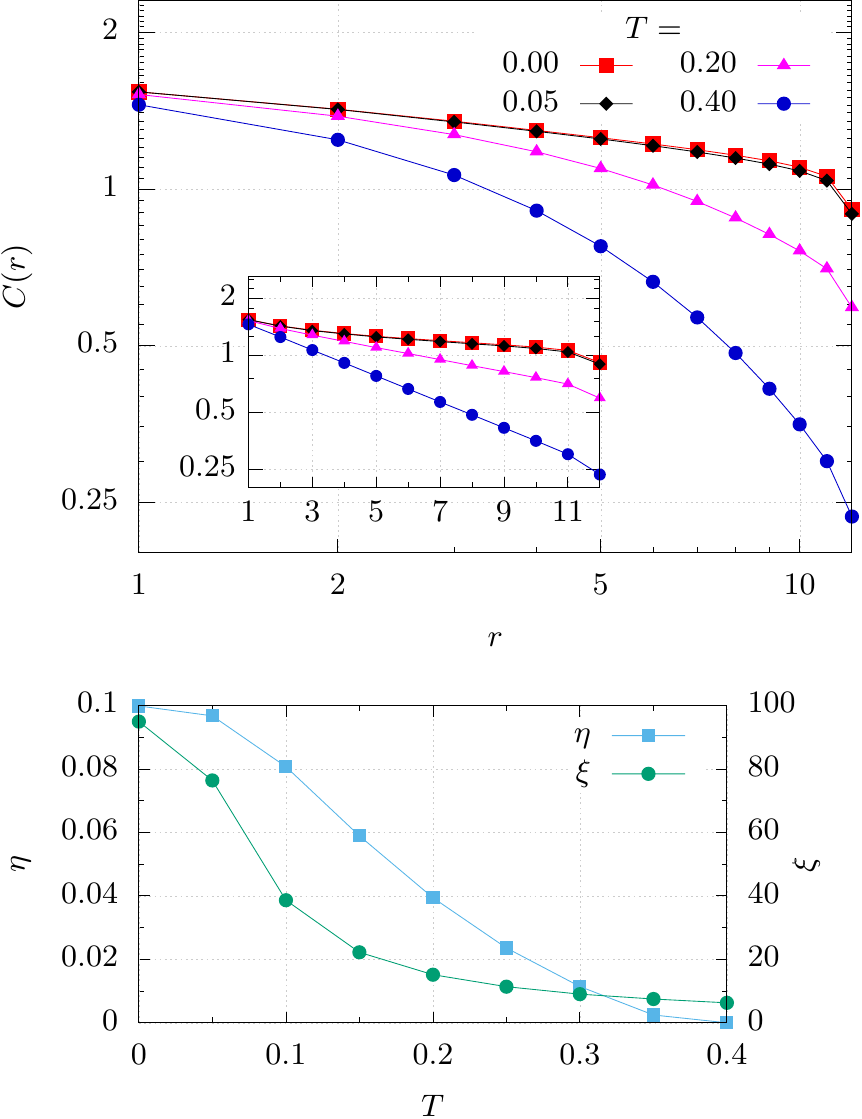}
	\caption{\label{fig:corrFunc}(Color online) Upper panel: Correlation functions $C(r)$ as a function of the distance $r$ in double-logarithmic scale (main plot), and semi-logarithmic scale (inset plot) for fixed~$J=0.46$, $U=1$, $\mu=1/2$ and various temperatures~$T$. The system size is $L=24$.
		Lower panel: Fit parameters~$\eta$ and~$\xi$, obtained by fitting Eq.~\eqref{eq:corrFit} to the correlation functions $C(r)$.}
\end{figure}

\subsection{\label{sub:superfluidChar}Characterization of superfluid features}

In order to identify superfluid features, we study the behavior of the two-point hopping correlation function $C(r) = \langle b^\dagger_j b^{}_{j+r} \rangle$. While in higher dimensions this correlation function exhibits long-range order in the superfluid phase, in one dimension the Mermin-Wagner theorem~\cite{Mermin1966, Hohenberg1967} prohibits a spontaneous breaking of the $\mathrm{U}(1)$ symmetry. Therefore, the superfluid phase merely exhibits quasi-long-range order in 1D, characterized by algebraically decaying correlation functions $C(r) \propto r^{-\eta}$. In contrast, outside of the superfluid phase order occurs only at a finite correlation length $\xi$, which is signaled by an exponential decay $C(r) \propto e^{-r/\xi}$. The quasi-long-range order can either be destroyed by thermal fluctuations, i.e.\@ when $T$ dominates over $J$, or, even at zero temperature, when $J$ becomes small and the particles crystallize due to density-density interactions, see the Mott insulator phase.

In order to illustrate this behavior, we plot the correlation functions $C(r) = \langle b^\dagger_{L/2} b^{}_{L/2+r} \rangle$ for a fixed value $J > J_c$ and different temperatures in Fig.~\ref{fig:corrFunc}. Although the finite size of the system (here $L=24$) makes it difficult to precisely extract the exponent $\eta$ or the correlation length~$\xi$, one can nevertheless detect the crossover from a power-law to an exponential decay when raising the temperature, hinting at a gradual loss of coherence. In order to quantify this observation, we fit the correlation functions with

\begin{equation}
	C(r) \propto \, r^{-\eta} \, \exp \left( -\frac{r}{\xi} \right) \; ,
	\label{eq:corrFit}
\end{equation}
and plot the fit parameters $\eta$ and $\xi$ as a function of~$T$ (see lower panel of Fig.~\ref{fig:corrFunc}): For small temperatures, we obtain $\xi \gg L$, meaning a predominantly algebraic decay, while for larger temperatures we have $\xi < L$ and $\eta$ small, signaling a mainly exponential decay.

\begin{figure}
	\includegraphics[width=0.95\columnwidth]{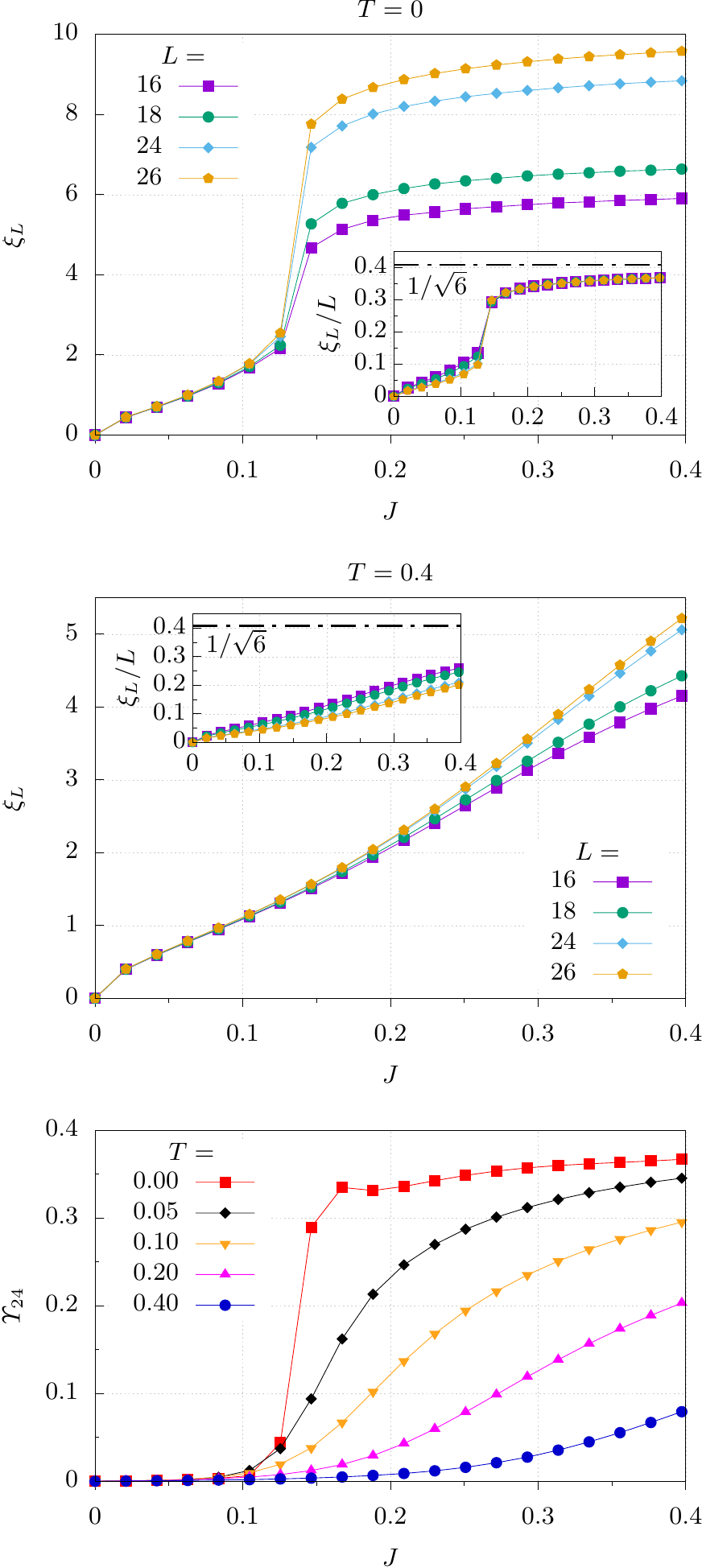}
	\caption{\label{fig:corrLen}(Color online) Finite-size correlation length $\xi_L$ as a function of~$J$ for several system sizes $L$ and two different temperatures $T=0$ (top panel) and $T=0.4$ (middle panel), with fixed $U=1$, $\mu=1/2$. The insets show $\xi_L/L$ for the same data, together with the upper bound $1/\sqrt{6}$. Bottom panel: Quantifier for superfluidity $\varUpsilon_{24}$, calculated with $\Delta L=2$ according to Eq.~\eqref{eq:superfluidQuant}, for various temperatures $T \in [0,0.4]$.}
\end{figure}

As a secondary approach to quantify the superfluid-like nature of a given state~$\rho$, we define and numerically calculate the ``finite-size correlation length''~$\xi_L$ as follows:
\begin{equation}
	\xi_L = \sqrt{ \frac{\sum_{j,k=1}^L  \left(j-k\right)^2 \langle b^\dagger_j b^{}_k \rangle}{\sum_{j,k=1}^L \langle b^\dagger_j b^{}_k \rangle} }\; .
	%\xi_L = \sqrt{ \frac{\sum\limits_{i=1}^L \sum\limits_{j=1}^L \left(i-j\right)^2 \langle b^\dagger_i b^{}_j \rangle}{\sum\limits_{i=1}^L \sum\limits_{j=1}^L \langle b^\dagger_i b^{}_j \rangle} }\; .
	%\xi_L = \sqrt{ \frac{\sum_{j=1}^L \left( j - L/2 \right)^2 \langle b^\dagger_{L/2} b^{}_j \rangle }{\sum_{j=1}^L \, \langle b^\dagger_{L/2} b^{}_j \rangle }  } \; .
	\label{eq:corrLen}
\end{equation}
The two definitions of $\xi$ coincide (neglecting a constant prefactor) when $\xi$ is larger than the lattice spacing, but smaller than the system size: $\xi_{L \gg \xi} = \xi$. If, however, the true correlation length becomes comparable to or larger than the system size, $\xi_L$ is upper-bound by a constant proportional to $L$. This bound can be shown by considering the limiting case of a constant correlation function $C(r) \to \varrho$, which is the asymptotically exact ground state correlation function of~$H$ in the limit $J\rightarrow\infty$, since $\eta(J{\to}\infty) \to 0$. Obviously, the true correlation length is diverging in this case ($\xi \rightarrow \infty$), but for $\xi_L$ we get
\begin{equation}
	\xi_L = \sqrt{ \frac{\sum_{j,k=1}^L \left(j-k\right)^2 \varrho}{\sum_{j,k=1}^L \varrho} }
	%\xi_L = \sqrt{ \frac{\sum\limits_{i=1}^L \sum\limits_{j=1}^L \left(i-j\right)^2 \varrho}{\sum\limits_{i=1}^L \sum\limits_{j=1}^L \varrho} }
	%\xi_L = \sqrt{ \frac{\sum_{j=1}^L \left( j - L/2 \right)^2 \varrho }{\sum_{j=1}^L \, \varrho } }
	= \sqrt{ \frac{L^2-1}{6}} \xrightarrow[L \to \infty]{} \frac{L}{\sqrt{6}} \; .
	\label{eq:corrLenJinf}
\end{equation}
More in general, one can show that for $L \to \infty$ the proportionality $\xi_L \propto L$ is valid for any algebraically decaying correlation function, if its exponent~$\eta$ is in the range $0\leq \eta < 1$. This condition holds throughout the superfluid phase~\cite{Giamarchi1997,Kuehner2000}. Consequently, a diverging correlation length~$\xi$ can be detected by monitoring whether the ratio~$\xi_L/L$ approaches a constant larger than zero when increasing $L$. If, on the other hand, this ratio tends to zero for increasing $L$, the correlation length is finite.

We illustrate this idea in Fig.~\ref{fig:corrLen} (upper two panels), both for zero and non-zero temperature. Clearly, the more superfluid-like the system, the more $\xi_L$ diverges with the system size~$L$. Based on this observation, we quantify the superfluid-like nature of a thermal state~$\rho$ via
\begin{equation}
	\varUpsilon_L(J,T) = \frac{\xi_{L+\Delta L}(J,T) - \xi_L(J,T)}{\Delta L} \; ,
	\label{eq:superfluidQuant}
\end{equation}
measuring incremental growth of $\xi_L$ while increasing the system size by~$\Delta L$. In the bottom panel of Fig.~\ref{fig:corrLen} we plot~$\varUpsilon_{24}$ with $\Delta L=2$ as a function of $J$ for different temperatures~$T$. At zero temperature, a sharp, discontinuous increase of~$\varUpsilon_{L}$ at $J\approx J_c$ separates the Mott insulating phase with vanishing~$\varUpsilon_L$ from the superfluid phase with non-zero~$\varUpsilon_{L}$. Higher temperatures gradually smooth out the transition and push the regime of superfluid-like correlations to larger and larger values of~$J$.

\subsection{\label{sub:phaseDiag}State diagram for finite system sizes at finite temperatures}

\begin{figure}
	\includegraphics[width=1.0\columnwidth]{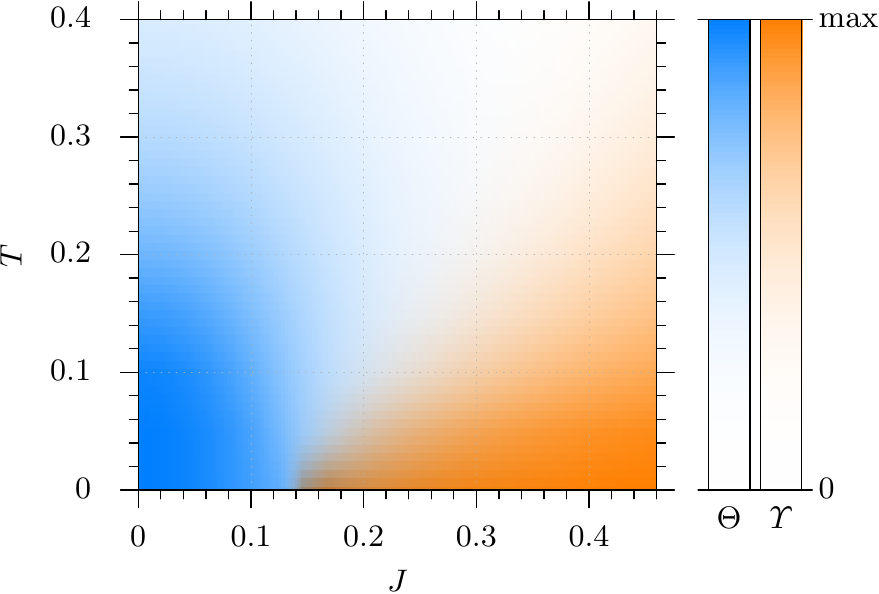}
	\caption{\label{fig:phaseDiag}(Color online) Characterization of Mott-like features (blue) and superfluid-like features (orange) as a function of hopping strength~$J$ and temperature~$T$ (for fixed $U=1$, $\mu=1/2$), based on the analysis described in Secs.~\ref{sub:mottChar} and~\ref{sub:superfluidChar}. White color corresponds to the ``thermal region''. The system size is $L=24$.}
	%and the chemical potential is $\mu = 1/2$
\end{figure}

Having developed quantifiers for both the Mott-like and the superfluid-like character of the system, we can summarize the data from the previous two subsections in a single graph, leading to the finite-size state diagram shown in Fig.~\ref{fig:phaseDiag}. The intensity of the blue color corresponds to the deviation $\Theta(J,T)$ of the variance $\sigma^2_{L/2}(J,T)$ from its maximal value in the considered intervals of $J$ and $T$. More specifically,
\begin{equation}
  \Theta(J,T) = \max_{J,T} \left[ \sigma^2_{L/2}(J,T) \right] - \sigma^2_{L/2}(J,T)
  \label{eq:mottinsQuant}
\end{equation}
with the variance~$\sigma^2_{L/2}(J,T)$ as defined in Eq.~\eqref{eq:variance}. Consequently, the intensity of the blue color encodes the presence of Mott-like features. Similarly, the intensity of the orange color encodes the occurrence of superfluid-like features measured via $\varUpsilon_{L}(J,T)$, as defined in Eq.~\eqref{eq:superfluidQuant}.

For $T=0$ the sharp transition between Mott insulator phase and superfluid phase at $J\approx J_c$ is clearly visible in Fig.~\ref{fig:phaseDiag}. For small enough temperatures and sufficiently far away from $J_c$ the essential features of the two phases survive. A larger and larger ``thermal region'', where thermal fluctuations prevent any type of order, opens up around $J_c$ when raising the temperature.

\section{\label{sec:dynamics}Dynamics}

We now discuss some aspects of time evolution, i.e.\ the out-of-equilibrium dynamics in the Bose--Hubbard model. In particular, we are interested in analyzing the behavior of the system when exposed to linear-ramp quenches in the hopping strength $J$ across the phase transition. This is the typical scenario investigated in the framework of the Kibble--Zurek mechanism (KZM)~\cite{Kibble1976,Zurek1985}. As before, we start from the zero temperature behavior and then proceed to analyze the impact of finite temperatures. We use the following quench protocol: 
\begin{enumerate}[(i)]
	\item
		The starting point is the equilibrium state $\rho_0$ of the Hamiltonian~$H$ for~$J=0$, $\mu=1/2$. Since the coupling term vanishes in this case, $\rho_0$ is always a product state. At zero temperature, $\rho_0$ is the pure state composed of the perfect Mott insulator state~$|\Psi\rangle$ with filling one, i.e.\@ $|\Psi \rangle = |1\rangle_1 \ldots |1\rangle_L$, while at finite temperature $\rho_0 = e^{-\beta H_0} / \, \text{Tr}[e^{-\beta H_0}]$.
		%the single-body density matrix contains non-zero populations of states with particle number larger than one.
	\item
		The initial state $\rho_0$ is evolved via unitary time evolution $\dot{\rho} = -i \, [H(t),\rho]$ in the time interval $t \in [-\tau_Q/2, \tau_Q/2]$, where $\tau_Q$ is the duration of the quench. The Hamiltonian is time-dependent through a linear ramp in the hopping strength
		\begin{equation}
			 J(t) = \frac{2 J_c}{\tau_Q} \, t + J_c \; ,
			 \label{eq:rampJ}
		\end{equation}
		which is chosen to be symmetric around the critical point~$J_c$, such that $J(0)=J_c$ and $J(-\tau_Q/2) = 0$. Since $[H(t), \sum_{j=1}^L n_j]=0$, the total particle number $N$ is a constant of motion. For $T=0$ this implies that the dynamics takes place along the line of constant filling~$\varrho=1$ in the phase diagram, which passes through the multicritical point at the tip of the first Mott lobe~\cite{Fisher1989}. The phase transition in this case~\footnote{Note that this is different from the scenario studied in Sec.~\ref{sec:static}, where not the filling~$\varrho$ but the chemical {potential~$\mu$} was kept constant} is of the BKT type~\cite{Berezinskii1972,Kosterlitz1973}, and it is located at $J_c\approx 0.30$~\cite{Krutitsky2016,Laflorencie2016}. The time evolution of the quantum many-body state (computed by means of MPS and LPTN for zero and finite temperature, respectively) is performed numerically with the Time-Evolving Block Decimation (TEBD)~\cite{Vidal2004} algorithm using a Suzuki-Trotter decomposition of the Hamiltonian at second order (see also Appendix~\ref{app:numerical_methods}).
	\item
		At the end of the quench, the superfluid correlation length~$\xi_\mathrm{fin}$ is measured using Eq.~\eqref{eq:corrLen}. We then study the behavior of this ``defect measure''~\cite{Zurek2005} as a function of the quench duration~$\tau_Q$.
\end{enumerate}

\subsection{\label{sub:quenchzero}Quenches at zero temperature}

In order to enable an understanding of the essential features of the system's state after the quench, the KZM provides a simple yet powerful argument relying on a comparison of the system's internal relaxation timescale~$\tau_R(t)$ with the external driving timescale~$\tau_D(t)$. This comparison separates the dynamics into two stages: an adiabatic stage when $\tau_R(t)<\tau_D(t)$, and an impulsed (sudden) stage when $\tau_R(t)>\tau_D(t)$. The instant~$\hat{t}$ at which the dynamics changes from adiabatic to sudden is called the ``freeze-out time''. Based on this simple picture, the KZM predicts that the order properties of the system after the quench are essentially determined by the instantaneous ground state at~$\hat{J} = J(\hat{t})$~\cite{Zurek1985}.

\begin{figure}
	\includegraphics[width=1\columnwidth]{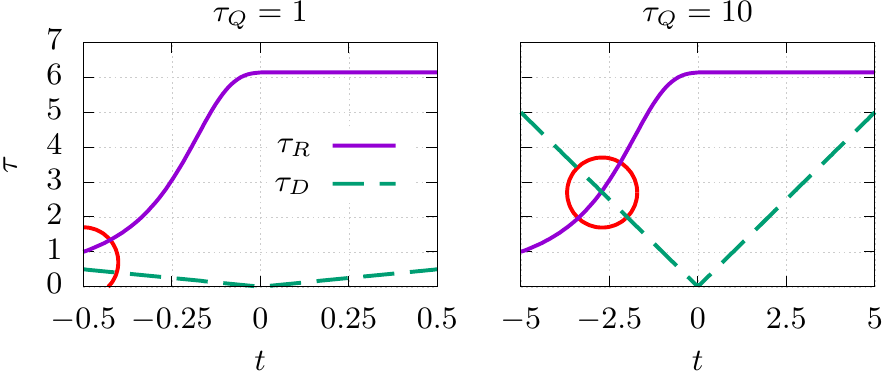}
	\caption{\label{fig:timescales}(Color online) Driving timescale~$\tau_D$ and relaxation timescale~$\tau_R$ as a function of time~$t$ for a system of $L=16$ sites undergoing the linear quench of Eq.~\eqref{eq:rampJ}, with fixed $U=1$, $\varrho=1$. Left panel: For~$\tau_Q < 2$ no intersection of the two timescales exists, hence~$\tau_D$ is always smaller than $\tau_R$. Right panel: For \mbox{$\tau_Q \ge 2$} the timescales intersect (depicted by a circle), leading to a nontrivial freeze-out time~$\hat{t}$.}
\end{figure}

For the case of a second order quantum phase transition, the KZM allows for a particularly elegant description of the scaling of the final density of defects as a function of the quench duration. More specifically, if at the critical point the equilibrium correlation length diverges with a critical exponent $\nu$ and the energy gap $\Delta E$ closes with another critical exponent $z\nu$, the KZM predicts~\cite{Zurek2005} that the final correlation length (after the quench) scales according to
\begin{equation}
	\xi_\mathrm{fin} \propto \tau_Q^\kappa \; , \quad \text{where} \quad \kappa = \frac{\nu}{1+z \nu}  \; ,
	\label{eq:KZscaling}
\end{equation}
i.e.\@ the scaling of the defect density as a function of the quench time is determined by a single constant exponent~$\kappa$.

\begin{figure}
	\includegraphics[width=0.95\columnwidth]{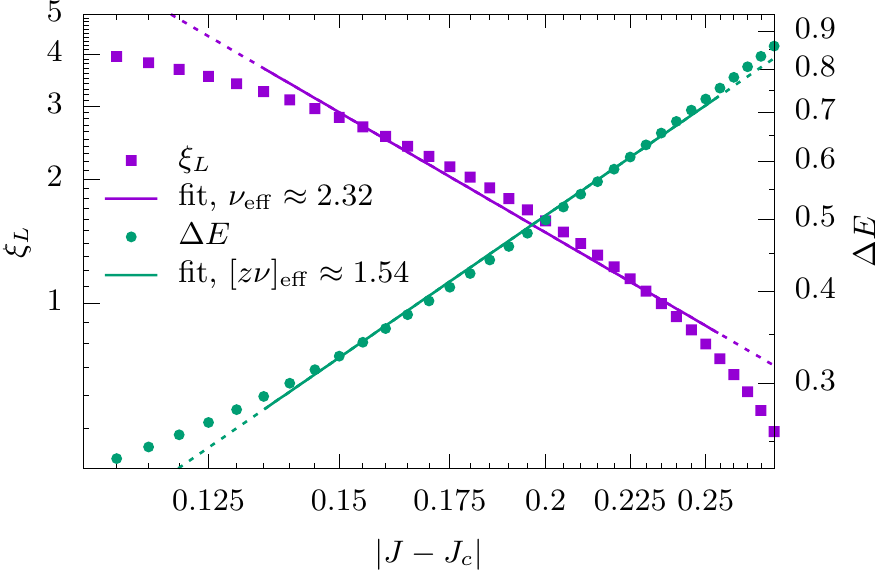}
	\caption{\label{fig:effectiveExponents}(Color online) Fitting of the effective critical exponents~$\nu_\mathrm{eff}$ and~$[z\nu]_\mathrm{eff}$ to the equilibrium scaling of~$\xi_L$ and~$\Delta E$, respectively, as a function of the distance $|J-J_c|$ from the critical point~$J_c$. Note that all axes are logarithmic. The fit interval (marked by non-dashed lines) has been chosen such that it covers the range of freeze-out points $\hat{J}$ for quench times in the interval $3\leq \tau_Q \leq 15$. The system size is $L=16$, and $U=1$, $\varrho=1$.}
\end{figure}

Here, however, due to the preservation of the total number of particles induced by the $\mathrm{U}(1)$ symmetry, we cross an infinite-order BKT transition which produces quantitative and qualitative deviations from the traditional KZ picture~\cite{Dziarmaga2014,Gardas2017}: While the basic idea of identifying the final correlation length with the one at equilibrium at time~$\hat{t}$ is in principle still valid, the exponential scaling~\cite{Kosterlitz1974} of the equilibrium quantities $\Delta E(J)$ and $\xi(J)$ near the critical point $J_c$ prevents the derivation of a simple expression like the one in Eq.~\eqref{eq:KZscaling}. Nevertheless, following Ref.~\cite{Dziarmaga2014}, one can still define ``effective'' critical exponents~$\nu_\mathrm{eff}$ and~$[z\nu]_\mathrm{eff}$ by approximating the exponentials with power-laws around a sufficiently small interval around the freeze-out point $\hat{J}$. Obviously, these exponents now depend on $\hat{J}$ and hence also on the quench time $\tau_Q$, but this approach allows one to recover (at least formally) the scaling given in Eq.~\eqref{eq:KZscaling}, after replacing $\kappa$ with an effective exponent~$\kappa(\tau_Q)$:
\begin{equation}
	\kappa(\tau_Q) = \frac{\nu_\mathrm{eff}(\tau_Q)}{1+ [z \nu]_\mathrm{eff} (\tau_Q)}  \; .
	\label{eq:effKZscaling}
\end{equation}
In the following, we will adopt this strategy to verify the validity of the KZM for the quench protocol described above at zero temperature. To this end, we first need to determine the freeze-out times $\hat{t}(\tau_Q)$. We do this numerically, by comparing the relaxation timescale $\tau_R(t) = 1/\Delta E(t)$ with the driving timescale $\tau_D(t) = |(J(t)-J_c)/\dot{J}(t)| = |t|$~\cite{Zurek2005}. This procedure is illustrated in Fig.~\ref{fig:timescales}, for two different quench times $\tau_Q$.
In our energy units we have $\Delta E (- \tau_Q / 2) = \Delta E (J = 0) = 1$, thus the relaxation timescale is always $\tau_R(-\tau_Q/2)=1$ at the beginning of the quench. Hence, for all $\tau_Q < 2$ the driving timescale $\tau_D(-\tau_Q/2) = \tau_Q/2$ is sufficiently fast that the quench will be completely sudden. On the other hand, if $\tau_Q \ge 2$, there is an intersection of the two timescales and a nontrivial KZM scaling of the final defect density can be expected. In order to verify the KZM in this regime, we numerically determine the effective critical exponents $\nu_\mathrm{eff}$ and $[z\nu]_\mathrm{eff}$ as shown in Fig.~\ref{fig:effectiveExponents}. By fitting power-laws to the equilibrium quantities in the appropriate interval of $J$, we obtain $\nu_\mathrm{eff} = 2.32 \pm 0.2$ and $[z\nu]_\mathrm{eff} = 1.54\pm 0.1$ which is compatible with the numbers reported in Ref.~\cite{Gardas2017}. Inserting these values into Eq.~\eqref{eq:effKZscaling} delivers the prediction $\kappa = 0.92\pm 0.12$ for the scaling of the final correlation length $\xi_\mathrm{fin}$ after the quench.

\begin{figure}
	\includegraphics[width=1.0\columnwidth]{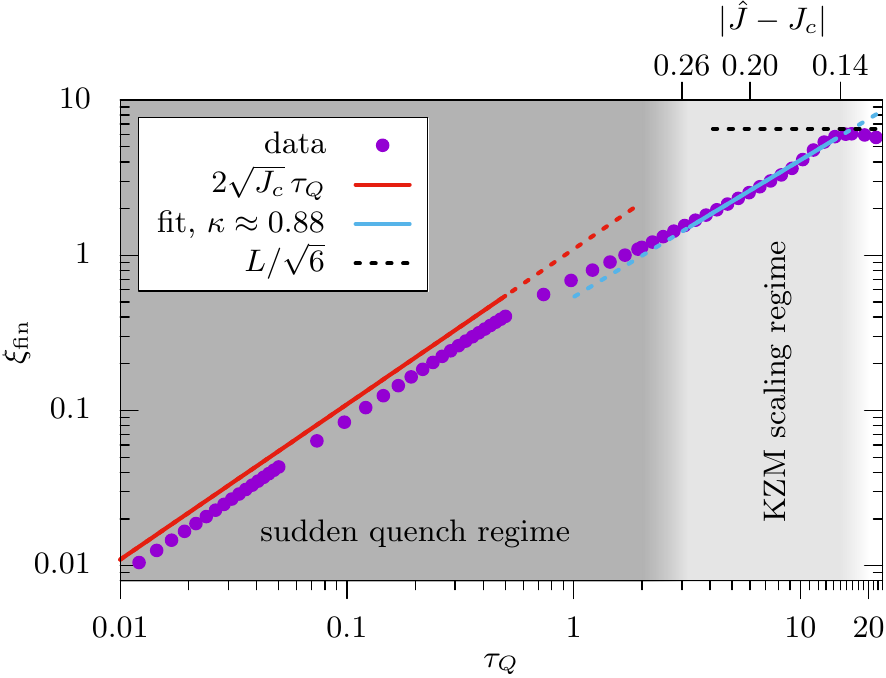}
	\caption{\label{fig:finalCorrZeroTemp}(Color online) Final correlation length~$\xi_\mathrm{fin}$ as a function of the quench time~$\tau_Q$ for~$T=0$. Three different regimes can be distinguished: sudden quench regime, KZM scaling regime, and the regime of finite-size saturation. The system size is~$L=16$, and $U=1$, $\varrho=1$.}
\end{figure}

In Fig.~\ref{fig:finalCorrZeroTemp}, we show the final correlation lengths $\xi_\mathrm{fin}$, measured after simulating the time evolution of the quantum many-body state with the TEBD algorithm, for various values of $\tau_Q$ spanning several orders of magnitude. Three regimes can be observed (marked by different shadings):
\begin{itemize}
	\item
		Sudden quench regime for $\tau_Q \lesssim 2$: As discussed above, the dynamics may be viewed as driven by a short impulse of duration~$\tau_Q$ in this regime. The fact that $\tau_Q$ is small allows for an approximate integration of the Schr\"odinger equation via discretization. Such an approximation can be done analytically, resulting in the following expression for the final correlation length (see Appendix~\ref{app:suddenQuench}):
		\begin{equation}
			\xi_\mathrm{fin}(\tau_Q) = 2 \sqrt{J_c} \, \tau_Q + \mathcal{O}(\tau_Q^2) \, .
			\label{eq:corrFinalSudden}
		\end{equation}
		For $\tau_Q \ll 1$ this expression is in good agreement with the numerical data, as demonstrated by the orange line in Fig.~\ref{fig:finalCorrZeroTemp}.
	\item 
		KZM scaling for $2  \lesssim \tau_Q  \lesssim 15$:  The fact that this regime has an upper bound for $\tau_Q$ is due to the finite size of the system (here $L=16$), implying a saturation value of $L/\sqrt{6}$ for the correlation length (see Eq.~\eqref{eq:corrLenJinf}). Fitting the exponent~$\kappa$ from the data in the KZM scaling regime yields $\kappa=0.88\pm 0.1$, which is in good agreement both with the prediction based on the equilibrium effective critical exponents outlined above and with the experimental and numerical results reported in Ref.~\cite{Braun2015}.
	\item 
		Saturated regime for $\tau_Q \gtrsim 15$: In this regime, the final correlation length is saturated due to the finite system size. Here, the defect density becomes too small to be resolved in a system of size~$L$ and the system appears completely ordered. Larger system sizes shift this regime to larger values of $\tau_Q$ (not shown in the figure).
\end{itemize} 

\subsection{\label{sub:quenchfinite}Quenches at finite temperatures}

In order to gain some (semi-quantitative) understanding of the behavior of the KZ scaling for finite temperatures, it is instructive to first consider the limiting case $T\to\infty$. Since in this case the thermal state $\rho_\mathrm{0}$ asymptotically approaches the identity, i.e.\ $\rho_\mathrm{0}(T{\to}\infty) \to \1$, the time evolution is trivial and the final state is again the identity: $\rho_\mathrm{fin}(T{\to}\infty) = \rho_\mathrm{0}$. This behavior automatically implies a vanishing KZ exponent~$\kappa$, because $\xi_\mathrm{fin}(\tau_Q) = \xi_\mathrm{0} = 0$. Hence, we expect $\kappa(T{\to}\infty) \to 0$.
On the other hand, for $T=0$ we need to recover the zero-temperature KZ exponent: $\kappa(T{\to}0) \to \kappa_0$, where~$\kappa_0$ is determined by the KZM described above. In order to provide a heuristic ansatz for the KZ exponent $\kappa(T)$ at finite temperatures, we resort to an Arrhenius argument. This argument fits because the deviation $\Delta \kappa(T) = \kappa_0 - \kappa(T)$ of the thermal KZ exponent from the zero-temperature KZ exponent can be viewed as a thermally induced quantity: An energy barrier needs to be overcome by means of thermal activation in order to enable an increase of~$\Delta\kappa$. Based on this motivation, we use the Arrhenius ansatz~\cite{Arrhenius1889}
\begin{equation}
	\Delta \kappa(T) = \kappa_0 \, e^{-E_a/T} \; ,
	\label{eq:arrhenius}
\end{equation}
where $E_a$ is an activation energy. Since the increase of~$\Delta \kappa(T)$ is ``activated" by an increasingly dominating population of excited states in the initial thermal density matrix~$\rho_0(T)$, the energy gap~$\Delta E^\prime$ between the ground and excited state can be considered an appropriate energy scale for the activation energy~$E_a$. As $\rho_0$ results from the Hamiltonian with $J=0$ in our protocol, the energy gap is easily seen to be $\Delta E^\prime=\mu$. It is worth mentioning that here the relevant energy gap is the inter-sector gap~$\Delta E^\prime$ (i.e.\ the energy difference between a system with~$N$ particles and a system with~$N+1$ particles) because we are working in a grand canonical ensemble. This gap is not the same as the intra-sector gap $\Delta E$ employed in the previous section, which is relevant for determining the freeze-out of the particle-conserving adiabatic time evolution of the ground state at zero temperature. 

\begin{figure}
	\includegraphics[width=1.0\columnwidth]{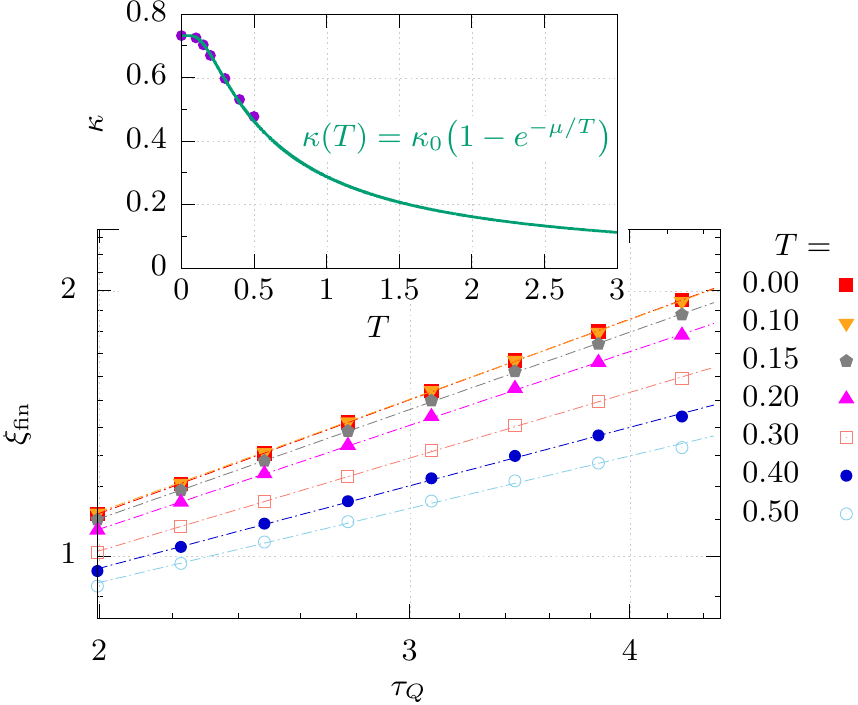}
	\caption{\label{fig:KZ_thermal}(Color online) Correlation length~$\xi_\mathrm{fin}$ after the quench as a function of the quench duration~$\tau_Q$, for various temperatures~$T$. The dashed lines are linear fits whose slopes determine the KZ exponents~$\kappa(T)$. The system size is~$L=16$, and $U=1$. Upper plot: extracted exponents as a function of~$T$, together with the Arrhenius ansatz indicated in Eq.~\eqref{eq:kappa_thermal}, with $\mu=1/2$.}
\end{figure}

Combining all assumptions, we predict the following thermal dependence of the KZ exponent:
\begin{equation}
	\kappa(T) = \kappa_0 \left( 1 - e^{-\mu/T} \right) \; .
	\label{eq:kappa_thermal}
\end{equation}
Figure~\ref{fig:KZ_thermal} stresses the validity of this ansatz: In the numerically accessible interval of small temperatures~($T\leq 0.5$), the determined KZ exponents follow indeed the prediction given in Eq.~\eqref{eq:kappa_thermal}.

\section{\label{sec:conclusion}Conclusion}

In this work, we have investigated the properties of the equilibrium and out-of-equilibrium one-dimensional Bose--Hubbard model at finite temperature. In the analysis of equilibrium properties we find, for the considered system sizes, a persistence of both insulating and superfluid features up to a certain temperature depending on the coupling $J$.
Our simulations yield a variety of observable data which characterize the physics of the thermal system. Additionally, theoretical predictions for the system's quantities at finite temperatures allow for thermometry in an experimental setup~\cite{Capogrosso2007,Trotzky2010}.

The investigation of the dynamical behavior of the system results in a verification of the Kibble--Zurek scaling for zero temperature, and a good agreement between the proposed Arrhenius-type ansatz and the obtained numerical data for $T>0$.

%Our study demonstrates the usefulness of TN tools for investigating the Bose--Hubbard model at finite temperatures, expanding the scope of this numerical method.

Our analysis offers many possible extensions, e.g.\ investigating certain regions of the $J\text{-}\mu$ phase diagram where a revival of the Mott insulating phase is expected~\cite{Kuehner2000} or, as often found to characterize experimental setups, simulating harmonically confined systems realized by site-dependent chemical potentials~\cite{Batrouni2002,Krutitsky2016,Frank2016}. 

Finally, an essential question is when the scenario of evolving a mixed state with a unitary time evolution following the von Neumann equation applies. Throughout this work we consider the case where the timescale of the quench is much shorter than the timescale of the system to reach the thermal equilibrium. This condition is normally fulfilled if we prepare the initial thermal state and are able to largely decouple the system from the environment. If, instead, the quench timescale is comparable to or larger than the relaxation timescale, one has to include open-system dynamics into the calculations~\cite{Yin2014,Yin2016a}. This case is left for future studies and requires a careful choice of Lindblad operators~\cite{Kossakowski1972,Lindblad1976,Gorini1976,BreuerPetruccione} for Markovian dynamics or evolution of non-Markovian systems~\cite{deVega2017,Mascarenhas2017}.

\begin{acknowledgments}
	We thank F. Tschirsich for discussions and additional numerical checks. Numerical calculations have been performed with the computational resources provided by the bwUniCluster project~\footnote{{b}wUniCluster: funded by the Ministry of Science, Research and Arts and the universities of the state of Baden-W{\"u}rttemberg,  Germany, within the framework program bwHPC}. We acknowledge financial support from the EU project QTFLAG, the BMBF via Q.com, and the Eliteprogramm for Postdocs of the Baden-W{\"u}rttemberg Stiftung via the TESLA.G project. S.M.\ gratefully acknowledges the support of the DFG via a Heisenberg fellowship and the TWITTER project.
\end{acknowledgments}

\appendix

\section{\label{app:numerical_methods}Numerical methods}

As mentioned above, we employ TN methods as our simulation tool. The ground state properties have been obtained via imaginary time evolution for MPS~\cite{Schollwock2011}, or via variational minimization for TTN~\cite{Tagliacozzo2009,Gerster2014}. In order to obtain thermal equilibrium states for $T>0$, we use imaginary time evolution applied to LPTN. Each discretization step of this evolution generates a fixed-temperature state, starting from the maximally mixed state (infinite temperature). The temperature of the LPTN after~$n$ steps is inversely proportional to~$n$, i.e.\ $T \propto 1/n$. In the following, we summarize the LPTN framework in more detail, beginning with a brief recap of MPS notation.

The idea of MPS is the decomposition of a
many-body wave function $\ket{\psi}$ representing a system on $L$ sites
into a set of $L$ local tensors which compose together $\ket{\psi}$.
The original vector representing the quantum many-body state has $d^{L}$ entries, where $d$ is the
local dimension of the Bose--Hubbard model in our case. Each tensor
represents one site and is of rank-3, $T_{\alpha_{j}, i_j,
\alpha_{j+1}}$; the index $i_{j}$ iterates over the different states
in the local Hilbert space, i.e.\ the Fock states on site~$j$, and
the indices~$\alpha$ connect the site to their nearest neighbor and
encode the entanglement to the complete subsystem on the left
and right, respectively. The maximal dimension of $\alpha$ enables
us to truncate entanglement and to keep simulations feasible; 
this maximal dimension is called bond dimension $m$. For $m = d^{L/2}$ the MPS representation covers all possible states, i.e.\ the full Hilbert space.

The idea of the LPTN lies in the positivity of a density matrix, i.e.,
we can decompose any density matrix $\rho$ as
\begin{eqnarray}
  \rho
  &=& U \Lambda U^{\dagger}
   =  U \sqrt{\Lambda} \sqrt{\Lambda} U^{\dagger}
   =  X X^{\dagger} \, ,
\end{eqnarray}
where $\Lambda_i \ge 0$ are the eigenvalues represented in a diagonal
matrix $\Lambda$, and we define $X \equiv U \sqrt{\Lambda}$. The matrix
$X$ is the purification of the density matrix $\rho$ and is sufficient
for the unitary time evolution and imaginary time evolution, as
outlined later on. The purification $X$ scales with the system size
as $d^{L} \times 1$ for a pure state and $d^{L} \times d^{L}$ for a
maximally mixed state $\rho \propto \1$. For the special case of
a pure state, there is exactly one eigenvalue equal to one and
$\rho = \ket{\psi} \bra{\psi}$. We can include this new index
running over the eigenvalues by extending each tensor in the MPS
with an additional index $\kappa_{j}$, i.e., $T_{\alpha_{j}, i_{j},
\kappa_{j}, \alpha_{j+1}}$; we obtain the LTPN. The MPS is regained
for $\mathrm{dim}(\kappa_{j}) = 1, \, \forall j$. In contrast, if
each $\mathrm{dim}(\kappa_{j}) = d, \, \forall j$, we regain, globally,
the dimension $d^{L}$ of the matrix $\Lambda$.

We turn to the argument why this representation is efficient in the
case of finite-temperature states. We define the thermal state as
$\rho_{\mathrm{th}} = \exp(- \beta H) / Z$ with the partition
function defined as $Z = \mathrm{Tr} \left[ \exp(- \beta H) \right]$. We can
rewrite the thermal state as
\begin{eqnarray}\label{eq:appa:itevo}
  \rho_{\mathrm{th}}
  &=& \frac{\exp(- \beta H)}{Z} \nonumber \\
   &=&  \frac{1}{Z} \exp\left(- \frac{\beta H}{2} \right)
      \1 \exp\left(- \frac{\beta H}{2} \right) ,
\end{eqnarray}
where the identity $\1$ is proportional to the infinite-temperature
state $\rho_{\mathrm{inf}}$; its purification can be
easily represented as an LPTN, where the global identity is
a product state of local identities with $\mathrm{dim}(\alpha_{j}) = 1, \, \forall j$;
the identity matrix is with respect to the indices $i_{j}$ and
$\kappa_{j}$ for each site~$j$. Equation~\eqref{eq:appa:itevo}
represents the imaginary time evolution with a constant
Hamiltonian. For the real-time evolution, we time-slice the Hamiltonian and
evolve the state under a Hamiltonian constant for each time step~$\Delta t$.
\begin{figure}
	\includegraphics[width=0.95\columnwidth]{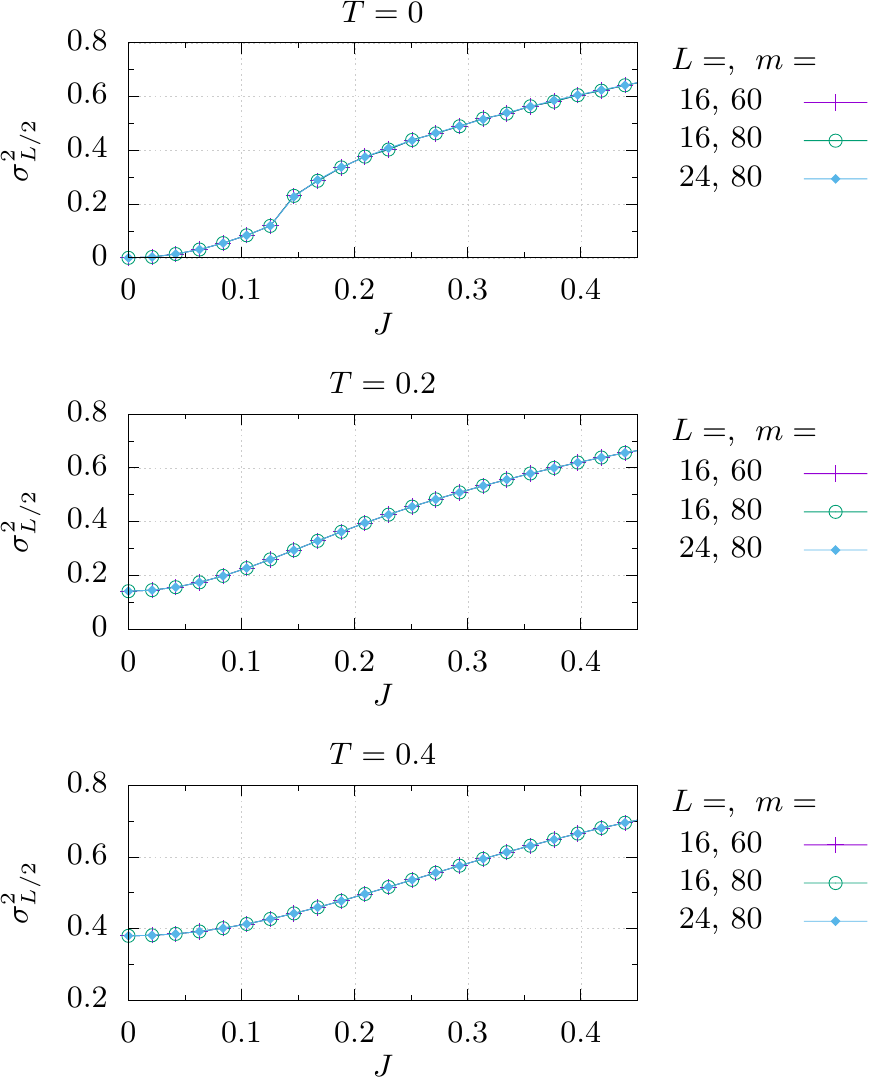}
	\caption{\label{fig:varianceConvergence}(Color online) Demonstration of convergence in the system size~$L$ and the bond dimension~$m$ for the variance in the middle of the chain as a function of the hopping strength~$J$. Local dimension $d=5$, and $U=1$, $\mu=1/2$.}
	\vspace{9mm}
	\includegraphics[width=0.95\columnwidth]{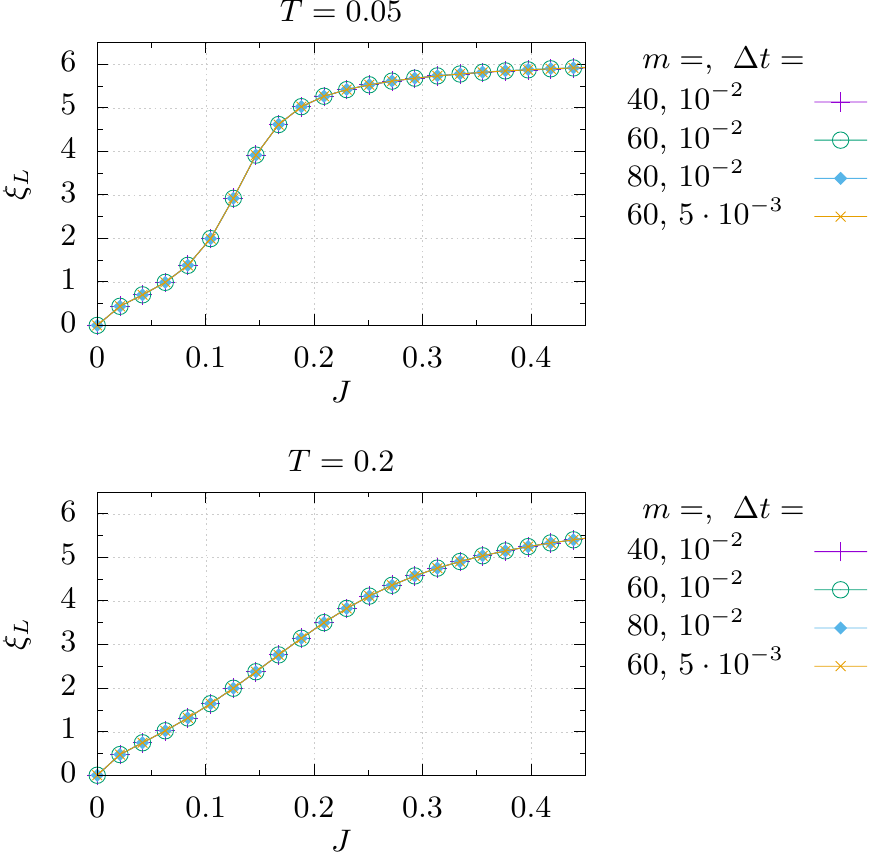}
	\caption{\label{fig:corrlenConvergence}(Color online) Demonstration of convergence in the bond dimension~$m$ and the Trotter time step~$\Delta t$ for the correlation length~$\xi_L$ as a function of the hopping strength~$J$. The system size is~$L=16$, the local dimension is $d=5$, and $U=1$, $\mu=1/2$.}
\end{figure}
To approximate the propagator of the Hamiltonian
in both time evolution schemes, we use a Suzuki-Trotter decomposition
splitting the Hamiltonian into $H = H_{2j - 1, 2j} + H_{2j, 2j+1}$,
where $H_{2j - 1, 2j}$ acts on odd sites and their nearest right 
neighbor, and the second term contains operators acting on even
sites and the nearest right neighbor. There is an error scaling with~$\Delta t$ when using
\begin{eqnarray}
  \exp(c H )
  &=& \exp\left( \frac{c}{2} H_{2j - 1, 2j} \right)
      \exp\left( c H_{2j, 2j + 1} \right) \nonumber \\
  &&  \times \exp\left( \frac{c}{2} H_{2j - 1, 2j} \right)
      + \mathcal{O}(\Delta t^3) \, ,
\end{eqnarray}
where the summands in each exponential on the right-hand side commute with each other and, consequently, can be exponentiated independently on the local two-site Hilbert spaces. The constant~$c$ is $i \Delta t$ and $\Delta t$ for the real
and the imaginary time evolution, respectively. The error of the total evolution
scales as $\mathcal{O}(\Delta t^2)$ for this second order Suzuki-Trotter
decomposition. The application of each of the three layers follows
the idea of the TEBD algorithm \cite{Vidal2004,Schollwock2011}. We
point out that it is sufficient to evolve either $X$ or $X^{\dagger}$
because both real and imaginary time evolution preserve the complex
conjugate structure of the purification.

In addition to errors scaling with the time step $\Delta t$, we
have to truncate correlations in the dimension of $\alpha_{j}$.
For an exact representation of a many-body state, the number of
weights at the center of our many-body representation can include
up to $d^{L}$ non-zero weights in the case of an LPTN, or $d^{L/2}$
for a pure state represented as an MPS. We allow for truncations
of small weights in the spectrum of the singular value decomposition (SVD).
The truncation with the SVD results in a minimal error for the
evolution of an LPTN~\cite{Werner2016} if the LPTN is properly
gauged~\cite{Silvi2017}. The index~$\kappa_{j}$ capturing the purification is not
growing during a unitary time evolution and guarantees an efficient
evolution.

\section{\label{app:convergence}Convergence of the simulations}

In this section we show with some examples that the numerical simulations presented above are at convergence with respect to changing the relevant refinement parameters. In our case, these include the system size~$L$, the bond dimension~$m$, the Trotter time step~$\Delta t$ and the local dimension~$d$. 

In Fig.~\ref{fig:varianceConvergence} we demonstrate that the variances of the particle occupations of the obtained equilibrium ground- and thermal states are independent of the system size~$L$. The deviations of $\sigma^2_{L/2}(J)$ are below point size when changing both the system size~$L$ and the bond dimension~$m$, for all considered values of the temperature~$T$. 

The two plots in Fig.~\ref{fig:corrlenConvergence} show that the Trotter time step~$\Delta t$ and the employed bond dimensions~$m$ are sufficient for converged imaginary time evolution results: The error of the correlation length~$\xi_L$ is again below point size.

\begin{figure}
	\includegraphics[width=0.9\columnwidth]{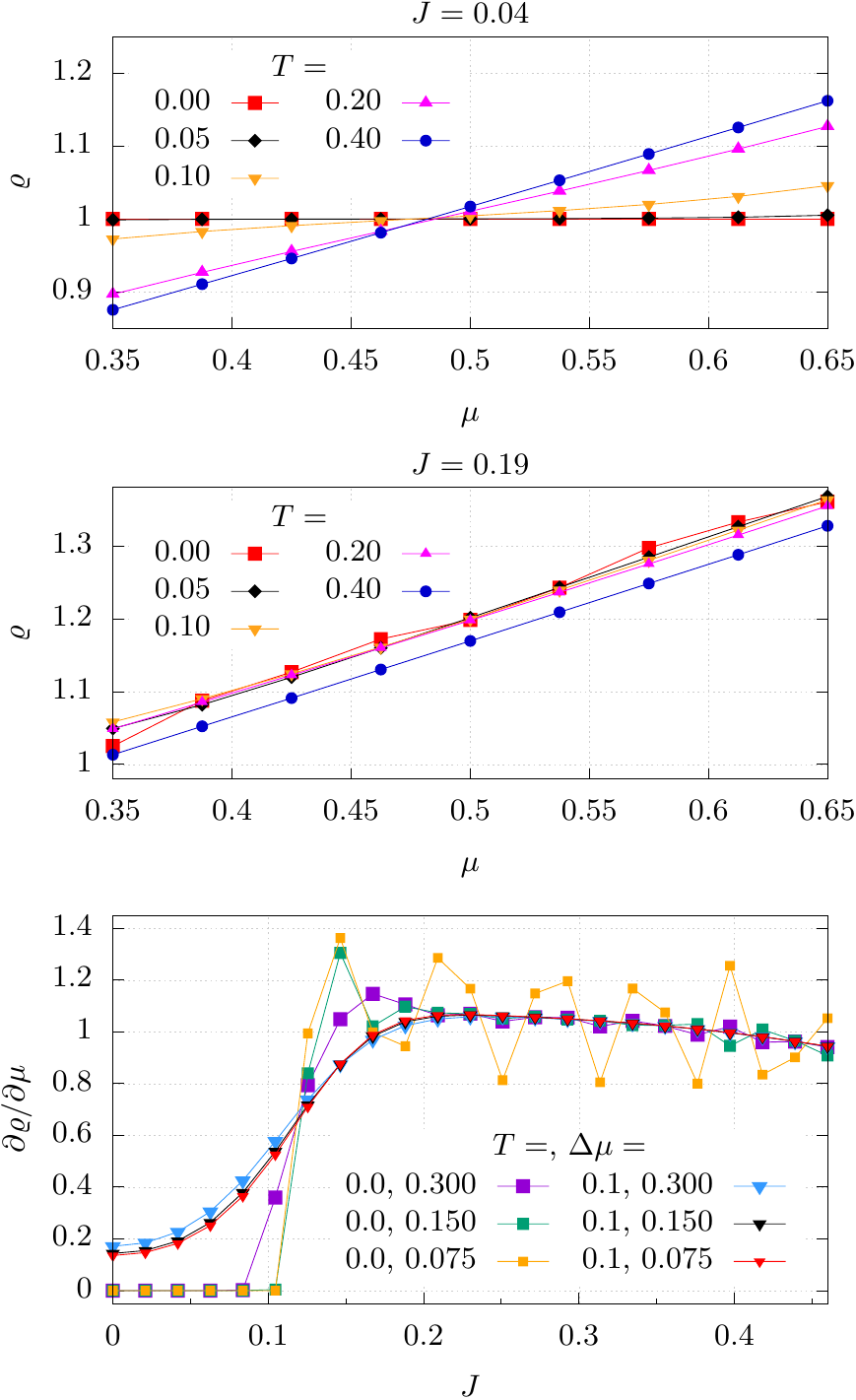}
	\caption{\label{fig:calcOfCompressibilily}(Color online) Numerical calculation of the compressibility~$\partial\varrho/\partial \mu$: The upper two panels show the filling~$\varrho$ as a function of the chemical potential~$\mu$, for various temperatures and two hopping strengths $J=0.04<J_c$ and $J=0.19>J_c$. The lower panel shows the fitted slopes~$\Delta\varrho/\Delta \mu$ as a function of the hopping strength~$J$, for two different temperatures and three different values for the fit interval~$\Delta\mu$. The system size is $L=18$, and $U=1$.}
\end{figure}

Figure~\ref{fig:calcOfCompressibilily} illustrates how we numerically obtain the compressibilities plotted in Fig.~\ref{fig:mottChar}: We first determine the filling~$\varrho$ as a function of the chemical potential~$\mu$, and then linearly fit this data in an interval of width~$\Delta\mu$ around~$\mu=1/2$, yielding an estimate for the compressibility $\partial\varrho / \partial\mu|_{\mu=1/2}$. We plot~$\varrho(\mu)$ for two different values of~$J$ in the upper two panels of Fig.~\ref{fig:calcOfCompressibilily}, demonstrating that $\partial\varrho / \partial\mu$ indeed only vanishes in the Mott insulator phase. For a system at zero temperature and finite size~$L$, the filling~$\varrho=N/L$ is limited to integer multiples of $1/L$, leading to a step-like behavior of $\varrho(\mu)$. In order to account for this finite-size effect, a careful choice of the fit interval $\Delta \mu$ is required. We find that $\Delta \mu=0.15$ provides a good trade-off in the parameter regime studied here, see lower panel of Fig.~\ref{fig:calcOfCompressibilily}.

\begin{figure}
	\includegraphics[width=0.95\columnwidth]{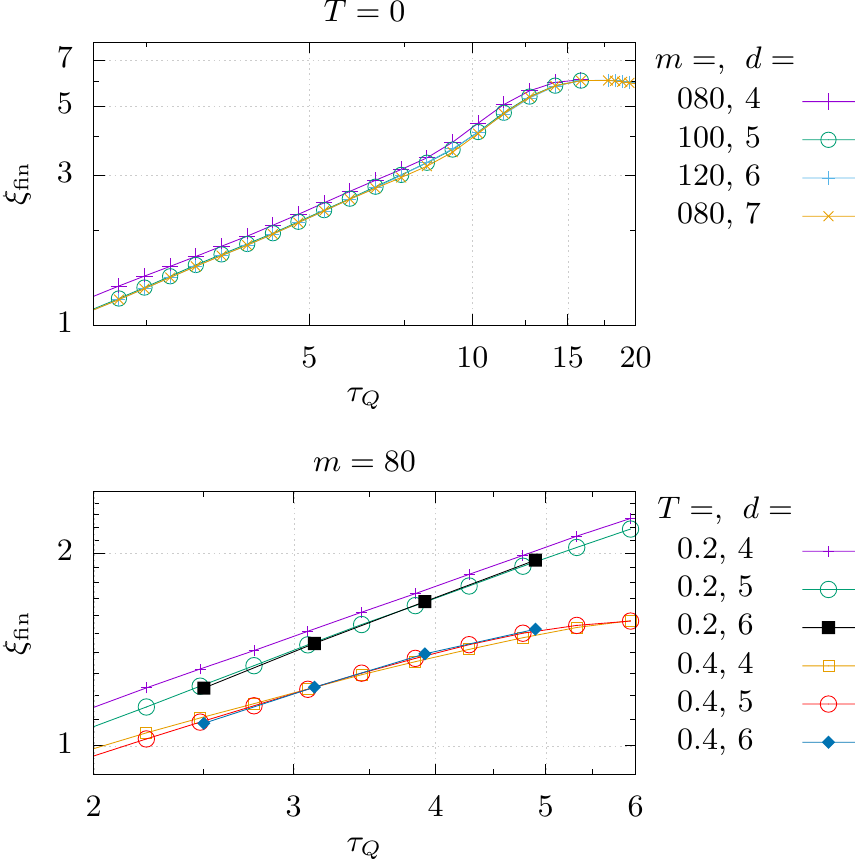}
	\caption{\label{fig:corrlenConvergenceDynamic}(Color online) Demonstration of convergence in the bond dimension~$m$ and local dimension $d$ for the final correlation length~$\xi_{\mathrm{fin}}$ as a function of the quench time~$\tau_Q$. The upper panel corresponds to the results presented in Fig.~\ref{fig:finalCorrZeroTemp}, the lower panel to the ones in Fig.~\ref{fig:KZ_thermal}. The system size is~$L=16$.}
\end{figure}

Finally, in Fig.~\ref{fig:corrlenConvergenceDynamic} we focus on the convergence of the quench data presented in Sec.~\ref{sec:dynamics}. It is evident that both for zero (upper panel) and finite temperature (lower panel) the local dimension~$d=4$ delivers noticeably different results compared to~$d=5$. Instead, the observed deviations between $d=5$ and $d\ge 6$ become negligibly small.\\
%Option: Mention independent check with Ferdinand's code.

\section{\label{app:suddenQuench}Analytical treatment of time evolution for short quenches}

Here we show the calculation leading to Eq.~\eqref{eq:corrFinalSudden}. We start from the perfect Mott insulator state with filling~$\varrho=1$, i.e.\@
\begin{equation}
	|\Psi_0 \rangle = |\Psi(-\tau_Q/2)\rangle =  |1\rangle_1 \ldots |1\rangle_L \; .
	\label{eq:psiInitial}
\end{equation}
Performing a linear quench in the hopping strength entails a non-trivial time evolution under the time-dependent Hamiltonian $H(t)$. We focus on the case $\tau_Q \ll 1$, meaning the total evolution time is short. Discretizing the integration of the Schr\"odinger equation we can write
\begin{equation}
	|\Psi(\tau_Q/2) \rangle \simeq \left(  1 - i \bar{H} \tau_Q - \\ \frac{1}{2} \bar{H}^2 \tau_Q^2 + \mathcal{O} \left( \tau_Q^3 \right) \right) | \Psi_0 \rangle \; ,
	\label{eq:psiFinalDisc}
\end{equation}
and we use the trapezoidal rule to determine the constant Hamiltonian~$\bar{H}$ during the discretization interval~$[-\tau_Q/2,\tau_Q/2]$:
\begin{equation}
	\bar{H} = \frac{1}{2} \left[ H(-\tau_Q/2) + H(\tau_Q/2) \right] = H(J(0)) = H(J_c) \; ,
	\label{eq:hamDisc}
\end{equation}
where we used the definition of the linear ramp $J(t)$ of Eq.~\eqref{eq:rampJ}. 

Plugging Eqs.~\eqref{eq:psiInitial} and~\eqref{eq:hamDisc} into Eq.~\eqref{eq:psiFinalDisc}, we can calculate the final state $|\Psi(\tau_Q/2)\rangle$, exact up to third order in~$\tau_Q$ (here we assume periodic boundary conditions for simplicity, and $U=1$):
\begin{widetext}
\begin{align}
	|\Psi(\tau_Q/2)\rangle 
	&= \left( 1 - 2 \,\tau_Q^2 J_c^2 L \right) |1\rangle_1 \ldots |1 \rangle_L \notag \\
	&+ \left( \frac{1}{\sqrt{2}} \, \tau_Q^2 J_c + i \sqrt{2} \, \tau_Q J_c \right) \sum_{j=1}^L |1\rangle_1 \ldots |1 \rangle_{j-1} \Big( |2\rangle_j | 0\rangle_{j+1} + |0\rangle_j |2\rangle_{j+1} \Big) |1\rangle_{j+2} \ldots |1\rangle_L \notag \\
	&- 2 \,\tau_Q^2 J_c^2 \sum_{j=1}^L \; \sum_{k>j+1}^L |1\rangle_1 \ldots |1 \rangle_{j-1} \Big( |2\rangle_j | 0\rangle_{j+1} + |0\rangle_j |2\rangle_{j+1} \Big) |1\rangle_{j+2} \ldots 
	|1\rangle_{k-1} \Big( |2\rangle_k | 0\rangle_{k+1} + |0\rangle_k |2\rangle_{k+1} \Big) |1\rangle_{k+2} \ldots |1\rangle_L \notag \\
	&- \frac{3}{\sqrt{2}} \, \tau_Q^2 J_c^2 \sum_{j=1}^L |1\rangle_1 \ldots |1 \rangle_{j-1} \Big( |2\rangle_j |1\rangle_{j+1} | 0\rangle_{j+2} + |0\rangle_j |1\rangle_{j+1} |2\rangle_{j+2} \Big) |1\rangle_{j+3} \ldots |1\rangle_L \notag \\
	&- \sqrt{6} \, \tau_Q^2 J_c^2 \sum_{j=1}^L |1\rangle_1 \ldots |1 \rangle_{j-1} |0\rangle_j |3\rangle_{j+1} | 0\rangle_{j+2} |1\rangle_{j+3} \ldots |1\rangle_L \notag \\
	&+ \mathcal{O}(\tau_Q^3) \; .
	\label{eq:psiFinalApprox}
\end{align}
\end{widetext}
From Eq.~\eqref{eq:psiFinalApprox} we obtain the two-site hopping correlations, again exact up to third order in $\tau_Q$:
\begin{equation}
	\langle b^\dagger_j b^{}_k \rangle = 
	\begin{cases}
		1, & \text{for } j=k \\
		2\,\tau_Q^2 J_c + \mathcal{O}(\tau_Q^3) \; , & \text{for } |j-k|=1 \\
		\mathcal{O}(\tau_Q^3), & \text{for } |j-k| > 1
	\end{cases}
	\label{eq:corrFinal}
\end{equation}
We can use these to determine the correlation length $\xi_\mathrm{fin}$ according to Eq.~\eqref{eq:corrLen}:
\begin{align}
	\xi_\mathrm{fin} &= \sqrt{\frac{ 2L \cdot 1^2 \cdot  2\,\tau_Q^2 J_c + \mathcal{O}(\tau_Q^3) }{ L + 2L \cdot 2\,\tau_Q^2 J_c + \mathcal{O}(\tau_Q^3)}}   \notag \\
	&= 2 \sqrt{J_c} \, \tau_Q + \mathcal{O}(\tau_Q^2) \; ,
	\label{eq:corrLenFinal}
\end{align}
which is the expression used in Eq.~\eqref{eq:corrFinalSudden}.

\bibliography{references}

\end{document}